\documentstyle[12pt]{article}

\begin{document}

\newcommand{\bib}{\bibitem}
\newcommand{\er}{\end{eqnarray}}
\newcommand{\br}{\begin{eqnarray}}
\newcommand{\be}{\begin{equation}}
\newcommand{\ee}{\end{equation}}
\newcommand{\epe}{\end{equation}}
\newcommand{\bea}{\begin{eqnarray}}
\newcommand{\eea}{\end{eqnarray}}
\newcommand{\ba}{\begin{eqnarray}}
\newcommand{\ea}{\end{eqnarray}}
\newcommand{\epa}{\end{eqnarray}}
\newcommand{\ar}{\rightarrow}
\def\I{{\cal I}}
\def\A{{\cal A}}
\def\F{{\cal F}}
\def\a{\alpha}
\def\b{\beta}
\def\r{\rho}
\def\D{\Delta}
\def\R{I\!\!R}
\def\l{\lambda}
\def\d{\delta}
\def\T{\tilde{T}}
\def\k{\kappa}
\def\t{\tau}
\def\f{\phi}
\def\p{\psi}
\def\z{\zeta}
\def\G{\Gamma}
\def\ep{\epsilon}
\def\hx{\widehat{\xi}}
\def\na{\nabla}
\newcommand{\bslash}{b\!\!\!/}
\newcommand{\vslash}{v\!\!\!/}
\newcommand{\eslash}{e\!\!\!/}
\newcommand{\rslash}{r\!\!\!/}
\begin{center}

{\large 2-Form Gauge Field Theories and ``No Go" for
Yang-Mills Relativistic Actions.}

\vspace{0.8cm}
 
Marcelo Botta Cantcheff\footnote{botta@cbpf.br, mbotta$\_$c@ictp.trieste.it}

\vspace{4mm} High Energy Section,

Abdus Salam ICTP, Strada Costiera 11, 34014 Trieste, Italy.

\vspace{3mm}

Centro Brasileiro de Pesquisas Fisicas (CBPF)

Coordena\c c\~ao de Campos e Particulas (CCP)

Rua Dr. Xavier Sigaud, 150 - Urca

22290-180 - Rio de Janeiro - RJ - Brazil \footnote{Permanent
Address.}.

\end{center}

\begin{abstract}
The transformation properties of a Kalb-Ramond field are those of
a gauge potential. However, it is not clear what is the group 
structure to which these transformations are associated. In this 
paper, we complete a program started in previous articles in order 
to clarify this question. Using the spectral theorem, we improve
and generalize previous approaches and find the possible group
structures underneath the 2-form gauge potential as extensions of
Lie groups, when its representations are assumed to act into any
tensor (or spinor) space with inner product.

We also obtain a fundamental representation where a two-form field
turns out to be a connection on a flat Euclidean basis manifold,
with a corresponding canonical curvature. However, we show that
these objects are not associated to space-time tensors and, in
particular, that a standard Yang-Mills action is not 
relativistically invariant, except (as expected) in the Abelian 
case. This is our main result, from the physical point of view.

\end{abstract}

\section{Introduction}

The (Abelian) Kalb-Ramond field \cite{kr} (KR), $b_{\mu\nu}$,
is a two-form field which appears in the low energy limit of
String Theory \cite{5it}, in Quantum Gravity \cite{6it}
and in several other frameworks in Particle Physics \cite{aplic}.
In particular, many attempts to incorporate mass to gauge fields 
in four dimensions take this object into account \cite{tm0,tm}. 
Its dynamics is governed by an action that is invariant under a 
symmetry transformation remarkably similar to that of a one-form 
gauge field \cite{la}: 
\be
b_{\mu\nu} \to b_{\mu\nu} + 
\partial_{[\mu}\b_{\nu]}, \label{trKR} 
\ee
where $\b_\nu$ is a 
1-form parameter. The question is: how can we associate the 
parameters $\b_{\mu}$ to the manifold of some gauge group 
\cite{otro,ultimo}? This problem was rigorously analyzed in
ref. \cite{gkr}, for specific spinor representations, and an 
Abelian connection was constructed with the KR field \cite{dkr}.

>From the physical point of view, it is essential to ask if {\it a
genuine} gauge theory may be formulated, {\it i.e}, if the 2-form
gauge potential may be stated as a connection on any group
manifold. This is important because, as is known, this structure 
would be crucial for the identities which determine the finiteness 
or not of physical models. In particular, in ref. \cite{h},
it has been proved that massive (non-Abelian) gauge models 
\cite{tm0,tm}, based on the KR field are ill-defined. The final
goal of this article is to argue that, in a very general
context, there are additional objections for non-Abelian gauge
actions with KR fields, that are associated with relativistic 
invariance. This kind of difficulties have already been indicated 
along different lines in previous literature \cite{teit}. Here, 
however, we analyze them from a very different perspective.

More recently, other points of view on the problem have been 
presented, related to minimal coupling with a matter field 
\cite{0it}, where expected applications in gravitation with 
torsion and Kalb-Ramond cosmology are pointed out 
\cite{0it,11it,12it,13it}.

On the other hand, it is well-known that anti-symmetric tensor 
fields, subjected to the gauge transformations (\ref{trKR}), are
equivalent to massless scalar particles \cite{{kr},5it,{tm},{1}}.
In a first-order formulation of this tensor gauge theory, it is
easy to show that there is an equivalence between these 
non-Abelian rank-2 anti-symmetric tensor fields, considered as 
gauge potentials, and torsionless non-linear $\sigma$-models 
\cite{town}. Recent models, that aim phenomenological applications 
\cite{helan}, explore this equivalence describing the 
interactions among vector mesons in terms of the geometrical 
properties of the target manifold. Spin-2 meson resonances may 
also be naturally accommodated, whenever the $\sigma$-model's 
target manifold is non-symmetric.

In this class of models, the rank-2 field plays the role of a 
torsion and its pull-back to space time allows the description of 
multiplets of excited spin-1 resonances. Another stream of 
research adopts the rank-2 gauge potential and explores its 
non-minimal coupling to charged matter to discuss the appearance
of defects, such as cosmic strings \cite{cris}, vortices 
\cite{alv} and magnetic monopoles \cite{win}.

This type of constructions, which defend the gauge character of 
the KR field, actually enforce the idea of a group structure since 
a closed algebra of infinitesimal variations is required.

This work is organized as follows: in Section 2, we deal with
the conditions to parametrize an Abelian group with an 1-form 
parameter. In Section 3, we extend this to the non-Abelian case. 
In Section 4, we recover a 2-form field from the connection and 
discuss the construction of gauge theories in this framework, 
pointing out the main conclusion.

\section{1-Form Group Parameter and Spectral Theorem}

Let us assume a four-dimensional flat space-time
$(M,\eta_{\mu \nu})$ and some Lie group, denoted by $G$, whose
associated algebra is ${\cal G}$; ${\tau}^{a}$ are the matrices
representing the generators of the group with $a = 1,\ldots ,
\mbox{dim}\:G$; $\tau_{abc}$ are the structure constants. Let us
take the gauge parameter to be an adjoint 1-form that can be
expanded as below: 
\be
\b_\mu = \b^a_\mu \tau^a . 
\ee
Consider also a vector space ${\cal S}=\{\psi \}$ where a 
representation for $G$ acts. We wish to find the transformation 
law under a group element parametrized by the object $\b^a_\mu$. 
Let us take an infinitesimal transformation parametrized by 
$\ep^a_\mu$. We shall get
\be
\psi'_I= (g_{\ep})_{IJ} (\psi )_{J} = ( I_{IJ} + i (M( \ep))_{IJ} 
+ (o^2(\ep))_{IJ} ) \psi_J \label{1ord}, 
\ee 
where, for an arbitrary 1-form $\b_\mu$, $M(\b)$ must be a linear 
operator from ${\cal S}$ into itself.

With the first order expression (\ref{1ord}), we may build a
group element corresponding to non-infinitesimal parameters, 
$\b_{\mu }$, by considering $\b^a_\mu= N\, \ep^a_\mu$ for a large 
integer $N$. Using the linearity of $M$, $M ( N \ep )=N M (\ep)$, 
we obtain $ g(\b)=\exp{i M(\b)} $.

For a better understanding of the problem, let us first consider
the case of an Abelian and unitary group ($G \sim 
U(1)$)\footnote{Later, we shall see that unitarity is related to 
the Euclidean character of space-time. Thus, in order to avoid 
this kind of question at this point, we assume a Riemannian 
space-time metric, since this does not represent any conceptual or 
technical inconvenient in field theories.}, the representation 
space ${\cal S}$ being assumed to be a complex vector space 
with an inner product. The matrix corresponding to the parameters 
$\b_{\mu }$ in the given representation, $M$, is a linear map from 
$\Lambda_1$ (the space of Lie algebra valued 1-forms) into the 
space of linear operators over ${\cal S}$ and 
\be
\d \psi= i M(\b)\psi \; .
\ee
By virtue of unitarity, $M$ is Hermitian, so 
that it can be diagonalized. Therefore, we can invoke the 
Spectral Theorem (ST), which states that, if $M$ has $K$ different 
eigenvalues, then there exist linear operators $E_1 , ..., E_K$ 
over ${\cal S}$, such that:
\begin{enumerate}
\item $ M(\b)= \sum\limits^K_{i=1} c_i(\b) E_i $, where $c_i$ are the eigenvalues of $M$.
\item $I= \sum\limits^K_{i=1}  E_i $.
\item $E_i E_j =0 $ , $ i \neq j$.
\item $E_i E_i = E_i$ ($E_i$ is a projector).
\item ${\cal S} = \bigoplus\limits_{i=1}^K {\cal S}_i $ (direct orthogonal sum), $E_i$ being the identity on the vector subspace ${\cal S}_i$.
\end{enumerate}
Notice that, although the eigenvalues $c_i$ are dependent on $\b$, 
the projectors $E_i$ are not, because, for an arbitrary pair of 
parameters $\b_1 ,\b_2$, $ [M(\b_1) , M(\b_2)]=0 $ (due to the 
Abelian character of the group), what makes them both diagonal in 
the same basis, thus defining the same set $\{E_i \}$.
On the other side, using that the $E_i$'s are $\b$-independent and 
satisfy 1-5, one can easily show that the commutator between 
$M(\b_1)$ and $M(\b_2)$ vanishes, for arbitrary $\b_1 ,\b_2$,
\be
\label{abalg}
[M(\b_1) , M(\b_2)] = \Sigma_{i,j} c_i(\b_1) c_j(\b_2) [E_i , E_j] = \Sigma_{i} c_i(\b_1) c_i(\b_2) (E_i - E_i)=0 \, .
\ee

Going on with our construction, as $M$ is a first-order 
homogeneous function of $\b$, then $c_i(\b)$ must be a real linear 
functional of the 1-form $\b$. Therefore, by definition,
there exist $K$ space-time vectors, $e_i$, such that $c_i = < \b ; 
e_i >$. Then, if we require that the full one form $\b$ can be 
recovered from the matrix $M(\b)$, {\it i.e.}, that the mapping 
$\b \to M(\b)$ is invertible, at least $d$ (the space-time 
dimension) of the $e_i$'s are linearly independent (otherwise it 
would be impossible to recover $\b$). So, we can verify that the 
dimension of the representation, $D$, satisfies $D\geq d$: it is 
obvious that $ D\geq K$; on the other hand, $\b $ recovery 
requires that $K \geq d$. Therefore, we obtain the first important 
conclusion on this group structure: the dimension of the 
representation has to satisfy the constraint $D\geq d$. We show 
below that, in fact, $D=d$ constitutes the minimal ({\it i.e.} 
fundamental) representation.

Notice, in addition, that each $E_i$ is associated to a vector 
$e_i$: let us take $\b=e_i$\footnote{Rigorously speaking, one 
should say: the 1-form (dual) corresponding to the vector $e_i$. 
However, we are relaxing, since we are considering an Euclidean 
metric for the space-time that is flat.}; then, since $c_i =1 , 
c_{j \neq i} =0$, we obtain that the projector must be given in 
terms of this unit vector.
\be
\label{EMi}
M(e_i)=E_i \;\; .
\ee
By virtue of this, notice that the set of diagonal matrices ${\cal 
M}$ is a closed algebra which can be decomposed in a direct sum of 
invariant spaces, each one associated with a single space-time 
direction:
\be
\label{dirsum}
{\cal M}= \bigoplus_i {\cal M}_i \; ,
\ee
which coincides with the main conclusion of ref. \cite{gkr}.

So, we can explicitly verify that a vectorial representation 
realizes this structure for $D=d=4$ and then constitutes a minimal 
representation.  Let us take as ${\cal S}$, the space of complex 
vectors tangent to the space-time base manifold, so that
the matrices $M$ are tensors  of type $(0,2)$. Due to the 
existence of a flat (Euclidean) inner product, $<\,;\,>$, we do 
not distinguish between vector and one-form, in our notation  
and, in what follows, we will not assume summation over repeated 
tensor indices.

Given an ordered orthonormal space-time basis $\{ e_\mu 
\}_{\mu=1}^4 $, the projectors are
\be
E_\mu = e_{\mu} \otimes 
e_\mu  ,\ee  and \be c_{\mu} (\b) = < \b \, ; \, e_{\mu} > .
\ee
Finally, we have the manifest (diagonal) form of the matrices representing the Lie Algebra  
\be
M(\b)= \sum\limits_{\mu =1}^{4} 
\; <\, \b \, ; \,  e_\mu \,> \; e_\mu \otimes e_\mu \, ,  
\ee
and, according to (\ref{EMi}), $ M(e_\mu)=E_\mu $ .

For a generic algebra valued 1-form parameter $\b = \b^a \tau^a $ 
(for an unitary group) the first part of the ST holds (1-5) but 
the commutation relations do not vanish. Then, we cannot see in 
this way that $E_\mu$ are independent of $\b$. We shall discuss 
the non-Abelian generalization later on.

Notice that if one takes another orthonormal space-time basis $\{ 
e'_\mu \}_{\mu=1}^4 $ related to the previous one by a Lorentz 
transformation\footnote{For us, an element of $SO(4)$, since we 
are assuming an Euclidean space-time metric.}
$e'_\mu = \sum\limits_{\nu } \;\lambda_\mu^{~~\nu} e_\nu$, then we must have, according to (\ref{EMi}):
\be 
\label{EM1}
M(e'_\mu)=E'_\mu \;\; .
\ee
On the other hand, the coefficients, referred to the previous basis 
are:
\be 
c_{\mu} (e'_\nu) = <  e'_\nu \, ; \, e_{\mu} > =  \sum\limits_{\a 
}\; \lambda_\nu^{~~\a}  < e_\a \, ; \, e_{\mu} >
=  \lambda_\nu^{~~\mu} \, ,
\ee
where we have used the orthonormality of the basis. Therefore, we 
get
\be
\label{M1}
M(e'_\mu)= \sum\limits_{\nu } \; \lambda_\mu^{~~\nu} E_\nu \;\; .
\ee
Comparing (\ref{EM1}) and (\ref{M1}), we obtain that the 
projectors in two different space-time cartesian coordinates
are related by:
\be
\label{EROT}
E'_\mu = \sum\limits_{\nu } \;\lambda_\mu^{~~\nu} E_\nu \;\; .
\ee

\section{Non-Abelian Extension.}

We can extend this structure to non-Abelian groups in a
natural way, simply by considering that the 1-form parameter takes
values in the algebra ${\cal G}$. So, the gauge parameter
reads $\b_\mu = \b^a_\mu \tau^a $ and the matrix corresponding to 
it is $M(\b)= \sum\limits_\mu \; \b_\mu  \; E_\mu$,  where $E_\mu$ 
are projectors ($E_\mu \, E_\nu = \delta_{\mu \nu} E_\nu $) 
satisfying the properties 2-5 of the spectral theorem. Thus, the 
commutation relation between two arbitrary algebra elements reads 
\bea
[M(\b) , M(\b')]= \sum_{\mu, \nu , a, b} \b^a_\mu \b^b_\nu [ 
E_\mu \tau^a ,   E_\nu \tau^b ] = \sum_{\mu, \nu , a, b} \b^a_\mu 
\b^b_\nu \delta_{\mu
\nu} E_{\nu} [\tau^a , \tau^b ] = \nonumber\\
=\sum_{\mu , a, b, c} \b^a_\mu \b'^b_\mu
E_\mu \tau^{abc}\tau^c = \sum_\mu [\b_\mu , \b'_\mu]\; E_\mu \; . 
\eea

It is straightforward to verify that the tensor products, $E_\mu 
\, \tau^a$, generate a well-defined algebra, that is an extension 
of the original Lie algebra (defined by the relations 
(\ref{subalg}) below). They obey the commutation relations:
\be
\label{algebra1}
[ E_\mu \tau^a ,  E_\nu \tau^b ] =\delta_{\mu
\nu} E_{\nu} [\tau^a , \tau^b ] 
\ee
\be
\label{algebra2}
[E_\mu \tau^a , \tau^b ]=E_\mu [\tau^a , \tau^b ]
\ee 
and, by definition, 
\be
\label{subalg}
[\tau^a , \tau^b ]= \tau^{abc} \, \tau^c \; .
\ee
They clearly satisfy Jacobi's identities (this can be seen by 
noticing that they are satisfied for the subalgebra (\ref{subalg}) 
and that the $E_\mu$ are projectors). After that, the construction 
of the gauge field theory (connection and field strength) can be 
pursued systematically.

Notice that, even in this non-Abelian generalization, the algebra 
obeys the structure shown by equation (\ref{dirsum}), being a 
direct sum of copies of the same group algebra for each space-time 
dimension. Therefore, the resulting group is the direct product of 
the  copies of $G$, ${\bf G} \sim \bigotimes\limits_\mu \; G_{(\mu)}$, as 
previously noticed in \cite{dkr}.

We can anticipate our final result, by observing that problems 
with relativistic invariance are already present in this 
non-abelian generalization of the algebra. If we notice that  
Lorentz  transformations are to be included in the automorphisms 
of this algebra, then the commutation relations (\ref{algebra1}) 
must be preserved:
\be
\label{algebra1rot}
[ E'_\mu \tau^a ,  E'_\nu \tau^b ] =\delta_{\mu
\nu} E'_{\nu} [\tau^a , \tau^b ] ,
\ee
where the generators
$E'_\mu$ and $E_\mu$ are related by (\ref{EROT}). Plugging 
(\ref{EROT}) into (\ref{algebra1rot}),
and using (\ref{algebra1}), we get
\be
\sum\limits_\a \lambda_\mu^{~~\a}  \lambda_\nu^{~~\a}   E_\a 
[\tau^a , \tau^b ] = \delta_{\mu\nu} \sum_\a  \lambda_\nu^{~~\a}   
E_\a [\tau^a , \tau^b ]\,.
\ee
If $[\tau^a , \tau^b ]\neq 0$ this implies that 
$\lambda_\mu^{~~\a}  \lambda_\nu^{~~\a} = 
\delta_{\mu\nu}\lambda_\nu^{~~\a} $,
which contradicts the fact that $\lambda$ is a generic homogeneous 
coordinate transformation. So, we conclude that, in the 
non-Abelian case, the set of algebra automorphisms does not 
contain the Lorentz group. Consequently, one has {\it different}
group structures for two arbitrary frames, which actually means 
that the group structure stated here is not relativistically 
invariant, except in the Abelian case.

Moreover, we are going to show explicitly that these problems 
remain at the level of objects with physical interpretation.

\section{Connection, Curvature and Gauge Theories}

Now, because of (\ref{abalg}), we have a well-defined group 
element acting on a complex space-time vector, $V^\alpha = 
v^\alpha + i w^\alpha \; \; v,w \in T_p M$, which reads as
\be 
g(\b)=\exp{i M(\b)} \; .
\ee
Therefore, using that $E_\mu$ are projectors ($E^n_\mu = E_\mu$ and $E_\mu E_\nu =0$, for $\mu \neq \nu$), a group element may be decomposed as
\be
\label{descg}
g(\b)=\exp{i M(\b)} = \sum_\mu \left( \sum_n \frac{i^n}{n!}
(\b_\mu)^n \right) \; E_\mu = \sum_\mu g_\mu (\b)
\; E_\mu 
\ee
where $g_\mu(\b)=\exp{i \b_\mu} $ and its inverse $
( g_\mu(\b))^{-1}=\exp{(- i \b_\mu )}$ are well defined
in a given Cartesian space-time basis. Obviously, $\b_\mu$ means 
{\it the $\mu$-component} of $\b$, which is a real number.

The covariant derivative acting on a vector field $V$ is written 
as $ \nabla_\mu \equiv \partial_\mu - i B_{\mu} $.
Thus, assuming that, under a gauge transformation $V \to V' = 
g(\b)V = e^{i M(\b)} V $, its covariant derivative transforms as
$\nabla'_\mu V'  = g(\b) \nabla_\mu V $,
and $\nabla'_\mu = \partial_\mu - i B'_{\mu}$, we see that the
connection must change according to 
\be
\label{transB}
B'_{\mu}= 
e^{i M(\b)}( B_{\mu} + i\partial_\mu M(\b)  )e^{-i M(\b) }.
\ee
Thanks to this property, we obtain that the infinitesimal 
transformation law for the connection reduces to the familiar 
expression:
\be
B'_{\mu} - B_{\mu} = \partial_\mu M(\b) + i [M(\b) , B_\mu ] + 
O(\b^2) \, ,
\ee
where $\partial_\mu M(\b) = \sum\limits_\nu \,(\partial_\mu \b_\nu 
\, ) \, E_\nu$,
%\Sigma_\nu \,(\partial_\mu \b_\mu \, ) \,e_\nu \; E_\nu \, ,\ee
as expected. This means that the manifest form of this connection 
is
\be
\label{Be}
B_\mu = \sum_\nu \, B_{\mu \nu} \,E_{\nu} \; \; ,
\ee
a $(0,2)$ type {\it tensor},  since it must be an object whose 
nature is preserved under gauge transformations.

Of course, this connection may be decomposed in its symmetric and
anti-symmetric parts, as $b_{\mu\nu}\equiv B_{[\mu\nu]}$ and
$G_{\mu\nu}\equiv B_{(\mu\nu)}$. We identify the (Abelian) 
Kalb-Ramond gauge field with $b_{\mu\nu}$.

Now, we define the field strength for the $B$-connection as usual:
\be 
[ \nabla_{\mu} , \nabla_{\nu} ] V = -i  F_{ \mu \nu} V \; ,
\ee
which results in
\be
F_{\mu \nu} = \sum_\r H_{\mu \nu \r} E_\r= 2\partial_{[\mu} 
B_{\nu]} + i [B_{\mu }, B_{\nu}]
=\sum_\r ( 2\partial_{[\mu} B_{\nu]\r} + i [B_{\mu \r}, B_{\nu\r}] 
)
E_\r .
\ee
Using the decomposition (\ref{descg}), it is straightforward to
verify the gauge covariance of the curvature tensor, 
\be 
F'= g(\b)F g(-\b) .
\ee
With this object we shall construct gauge invariant actions. In
doing this, let us consider
\be 
F  F = \sum_\r E_\r \; H_{\mu \nu \r} H_{\a \b \r}
\ee
which is also gauge covariant $F'  F'= g(\b) F  F g(-\b)$.

Notice that this curvature is not a space-time tensor
(except in the Abelian case) since we cannot associate a tensor to 
both terms simultaneously.

Consider a generic (diagonal) matrix $ C = \sum\limits_\mu c_\mu 
\; E_\mu $, such that $c$ is any $(0,p+1)$-tensor and each 
component $c_\mu$, with respect to the basis $\{e_\mu\}$, denotes 
a generic $(0,p)$-tensor. It may be thought as a map
$C: {\cal S} \times {\cal S}^* \to \Pi_p $ ($\Pi_p$ denotes the 
space of tensors of type $(0,p)$)\footnote{Since we are
considering an Euclidean flat metric, there is a trivial 
identification between ${\cal S}$ and its dual ${\cal
S}^*$.}. Then we construct the basis dependent vector
$ {\bf e} \equiv \sum\limits_\mu e_\mu \in {\cal S} $. In this 
way, we can recover the full tensor $c$, by multiplying ${\bf e}$ 
and $C$,
\be
{\bf e} \;  C = c \; ,
\ee
which is a $(0, p+1)$-tensor. Therefore, it is easy to verify 
that, multiplying the curvature by ${\bf e}$, we obtain:
\be
{\bf e}\; F = 2\partial_{[\mu} B_{\nu]\r } + i\sum_{\r} 
[B_{\mu \r}, B_{\nu\r}]\; . 
\ee
In the above expression, we can express the second term as
\be
{\bf e}\; F = 2\partial_{[\mu} B_{\nu]\r } + i\sum_{\a \b} [B_{\mu 
\a},
B_{\nu\b}] \delta_{\a \b}\; .
\ee
We manifestly observe that the tensorial type of both terms are
different (a $(0,3)$-tensor plus a $(0,2)$-tensor). Then, clearly,
not only this is not a tensor but also there is no algebraic
function of it which is. In particular, if we define a
Yang-Mills action as any (gauge invariant) {\it algebraic
functional} of the curvature tensor, we conclude that it does not
exist. This is the central argument of the ``no go" statement
claimed above.

A kind of Yang-Mills action on an Euclidean (four dimensional) 
space-time should be 
\be
S_{\rm YM}\equiv \int d^{4}x
g^{\mu \nu \a \b}\; {\rm Tr}_{(\rm algebra)}\, F_{\mu \a } \, F_{\nu  
\b},
\ee
where $g^{\mu \nu \a \b}$ is some object that does not depend on
the gauge fields. In particular, the standard Yang-Mills model
corresponds (in Euclidean space-time) to $g^{\mu \nu \a
\b}= \delta^{\mu \nu} \delta^{\a \b}$. Again, the price of keeping
Lorentz invariance of the theory is that Yang-Mills-like gauge
actions cannot be defined in this framework.

In the Abelian case, however, this is  a well-defined Lorentz 
invariant gauge theory, whose underneath group structure has been 
established here. So, in the fundamental (vector) representation 
${\rm tr} E_\mu = < e_\mu ; e_\mu >= 1$ and this may be expressed 
as:
\be
S_{\rm Abelian}[B_{\mu \nu }] \equiv \int d^4x {\rm tr}( F \; 
F )\; ,
\ee
which is a gauge invariant Abelian action.

{\bf Aknowledgements}: The author is indebted to
J. A. Helay\"el-Neto, A. Lahiri and G. Thompson for useful comments.
S. Alves Dias is specially aknowledged for reading the manuscript and
for helpful criticisms.
 Thanks are also due
to the referee of this paper for pointing out central aspects of 
the relativistic invariance of the non-Abelian structure and for 
calling my attention to fundamental references on this subject.
The author is fellowed by CLAF/CNPq.

\end{document}